\documentclass[preprint, aps,prb,amsmath,amssymb,showpacs,superscriptaddress,seceqn]{revtex4-1}
\usepackage{amsfonts, hyperref, times}
\usepackage{graphicx}
\usepackage{graphics}
\usepackage{bbold}
\usepackage{siunitx}

\usepackage{ragged2e}
\usepackage{caption}
\providecommand{\vect}[1]{{\boldsymbol{#1}}}

\begin{document}
  
\title{A magnetic skyrmion as a non-linear resistive element -- a potential building block for reservoir computing}

\author{Diana Prychynenko\thanks{prychynenko@uni-mainz.de}}
\email[Diana Prychynenko is the corresponding author for discussions related to skyrmion physics: ]{prychynenko@uni-mainz.de}
\affiliation{\footnotesize Institute of Physics, Johannes Gutenberg University Mainz, 55099 Mainz, Germany}
\affiliation{\footnotesize Graduate School of Excellence Material Science in Mainz, 55128 Mainz, Germany}

\author{Matthias Sitte}
\affiliation{\footnotesize Institute of Physics, Johannes Gutenberg University Mainz, 55099 Mainz, Germany}

\author{Kai Litzius}
\affiliation{\footnotesize Institute of Physics, Johannes Gutenberg University Mainz, 55099 Mainz, Germany}
\affiliation{\footnotesize Graduate School of Excellence Material Science in Mainz, 55128 Mainz, Germany}
\affiliation{\footnotesize Max Planck Institute for Intelligent Systems, 70569 Stuttgart, Germany}

\author{Benjamin Kr{\"u}ger}
\affiliation{\footnotesize Institute for Laser Technologies in Medicine and Metrology, University of Ulm, 89081 Ulm, Germany}

\author{George Bourianoff}
\email[George Bourianoff is the corresponding author for discussions related to reservoir computing: ]{george.bourianoff@intel.com}
\affiliation{\footnotesize Intel Corporation, 1300 S. MoPac Exp, Austin, Tx, 78746, U.S.}

\author{Mathias Kl{\"a}ui}
\affiliation{\footnotesize Institute of Physics, Johannes Gutenberg University Mainz, 55099 Mainz, Germany}

\author{Jairo Sinova}
\affiliation{\footnotesize Institute of Physics, Johannes Gutenberg University Mainz, 55099 Mainz, Germany}
\affiliation{\footnotesize Institute of Physics ASCR, v.v.i, Cukrovarnicka 10, 162 00 Praha 6, Czech Republic}

\author{Karin Everschor-Sitte}
\affiliation{\footnotesize Institute of Physics, Johannes Gutenberg University Mainz, 55099 Mainz, Germany}

\date{\today}

\begin{abstract}
Inspired by the human brain, there is a strong effort to find alternative models of information processing capable of imitating 
the high energy efficiency of neuromorphic information processing. One possible realization of cognitive computing are reservoir computing networks. These networks are built out of non-linear resistive elements which are recursively connected.
We propose that a skyrmion network embedded in frustrated magnetic films may provide a suitable physical implementation for reservoir computing applications. The significant key ingredient of such a network is a two-terminal device with non-linear voltage characteristics originating from single-layer magnetoresistive effects, like the anisotropic magnetoresistance or the recently discovered non-collinear magnetoresistance.
The most basic element for a reservoir computing network built from ``skyrmion fabrics'' is a single skyrmion embedded in a ferromagnetic ribbon. 
In order to pave the way towards reservoir computing systems based on skyrmion fabrics, here we simulate and analyze 
i) the current flow through a single magnetic skyrmion due to the anisotropic magneto-resistive effect and 
ii) the combined physics of local pinning and the anisotropic magneto-resistive effect.
\end{abstract}

\pacs{}

\maketitle

\section{Introduction}
\subsection{Reservoir computing}
Present day computers are primarily based on advanced CMOS transistors executing Boolean logic functions and as such are subject to fundamental tradeoffs between energy efficiency and speed.  This has led to intense world-wide research on alternative modes of computing. 
There is a strong effort to find energy efficient models of information processing which are inspired by biological neural networks. These approaches try to imitate neuromorphic information processing models.\cite{Gershman2015} While CMOS devices appear to be optimal for Boolean logic, implementing alternative models of computation implies using alternative physical systems beyond CMOS. Several approaches to neuromorphic computing are being explored.
One example is the concept of reservoir computing (RC).\cite{Burger2015, Lukosevicius2012, Verstraeten2007, Maass2002}
It has long been recognized\cite{Legenstein2005} that some aspects of human cognition could be modeled by recursively connected dynamical systems composed of randomly connected non-linear devices. Currently, such systems have been realized using fabrics of self-assembled memristive devices.\cite{Wang2009b, Crutchfield2010, Avizienis2012, Hasegawa2010, TingChang2013, Sillin2013, Stieg2012, Wang2015f, Huang2016b, Lequeux2016} 
The complexity of these networks resembles that of biological brains, in which the complex morphology and interactions between heterogeneous and non-linear network elements are responsible for powerful and energy-efficient information processing\cite{Sporns2016}. Unlike designed computation,  where each device has a specified role, computation in random resistive switch networks does not rely on specific devices in particular roles. Instead, it is encoded in the collective non-linear dynamic mapping behavior of a network on an applied input signal.

Previous research has described reservoirs implemented with memristors, quantum dot lasers, and atomic switches.\cite{Demis2015}
Here, we propose to use magnetic textures, more precisely skyrmions, to build a new type of RC system which has not been suggested yet, but offers potential advantages with respect to size, efficiency and complexity.

\subsection{Magnetic skyrmions}

Magnetic skyrmions\cite{Muhlbauer2009a} are chiral spin structures that can be characterized by a topological index, the winding number.\cite{Skyrme1962, Bogdanov1989, Nagaosa2013}.
Because of their topology they contain quantized amounts of magnetic flux and cannot be unwound easily\cite{Milde2013, Nagaosa2013}
making them robust to defects and impurities.
\cite{Rosch2013, Mueller2015} They can be created,\cite{Romming2013,Jiang2015a, Everschor-Sitte2016, Hrabec2016} read, moved, \cite{Schulz2012, Jonietz2010, Electrodynamics2012, Iwasaki2013, Sampaio2013}   and excited\cite{Schwarze2015} via magnetic and electric fields,\cite{Mochizuki2012, White2012, Ogawa2015}  spin currents and magnons.\cite{Schutte2014, Iwasaki2014, Schwarze2015, Petrova2011a}

Skyrmions can occur due to various mechanisms, which can be brought down to the interplay of interactions favoring different alignments of the magnetization. As such they occur in several systems (including metals, semiconductors and insulators)\cite{Yu2010, Pfleiderer2010} and at different sizes (as small as $\SI{1}{\nano\metre}$).\cite{Heinze2011, Romming2013} In particular, they also exist in technologically relevant material systems  at room temperature.\cite{Woo2016, Yu2011b, Gilbert2015}

Skyrmions occur mostly as either Bloch or N\'eel type depending on the details of the antisymmetric exchange interaction. 
The profile of a Bloch skyrmion is a spiral (spin rotates in the tangential planes) and for a N\'eel skyrmion it is a cycloidal (spin rotates in the radial planes). 
There are currently intensive research studies to exploit the full potential of magnetic skyrmions in spintronics applications, such as ultra-high density information storage and information processing.\cite{Fert2013, Tomasello2014, Zhang2015c, Muller2016}
Proposals to realize ``classical computing'' for example by building skyrmion based logic gates\cite{Zhang2015} and transistors\cite{Zhang2015b} have been put forward.

\subsection{Magnetic skyrmions for reservoir computing}

Recent simulations have shown spontaneous, highly complex, and self-organized patterns of skyrmions\cite{You2015} in thin film configurations which are very well suited for RC. 
The basic element of such a network is an isolated skyrmion in a ferromagnetic ribbon.  For RC, the property of the single skyrmion to cause a magnetization dependent magnetoresistive effects is crucial. We stress that the particular type of the magnetoresistance (MR) is not important and that the results of our analysis regarding RC can be transferred. Examples for magnetization dependent magnetoresistive effects are the anisotropic magnetoresistance (AMR) as well as the recently discovered non-collinear magnetoresistive (NCMR) effect\cite{Hanneken2015, Kubetzka2017} which has a much larger MR change. 

Here, we analyze the contribution of the (non-crystalline) anisotropic magnetoresistance (AMR) originating from magnetic skyrmions.
In ferromagnetic samples, the resistance depends on the relative orientation of the magnetization and the direction of the electric current.\cite{McGuire1975} 
The AMR effect is caused by anisotropic scattering of the charge carriers in the presence of spin-orbit coupling. The resultant resistivity is commonly larger when the current and the magnetization are parallel, and smaller when they are perpendicular.
This leads to a distinct contribution of confined spin structures to the resistance as previously demonstrated for domain walls.\cite{Klaui2003}
With this work we pave the way towards skyrmion fabrics based RC.

\subsection{Structure of the paper}
In Sec.~\ref{model}, we first describe our approach and then we show in Sec.~\ref{results1} our numerical results for the current distribution of single skyrmions (Bloch and N\'eel type) embedded in a ferromagnetic background based on the AMR effect. 
In this section we also analyze the current-voltage characteristics of an unpinned moving skyrmion in the conducting channel. 
In Sec.~\ref{results2}, we consider in a second step  the non-linear current-voltage characteristics of a single skyrmion that is pinned by a local change in the anisotropy.
In Sec.~\ref{sec:pathtoRCwithskrymions}, we outline the path to RC with skyrmion fabrics and substantiate that our results obtained in this work provide a key step in reaching this goal.

\section{Our model}
\label{model}

\subsection{Geometry} 
\begin{figure}
   \includegraphics[width=0.5\textwidth]{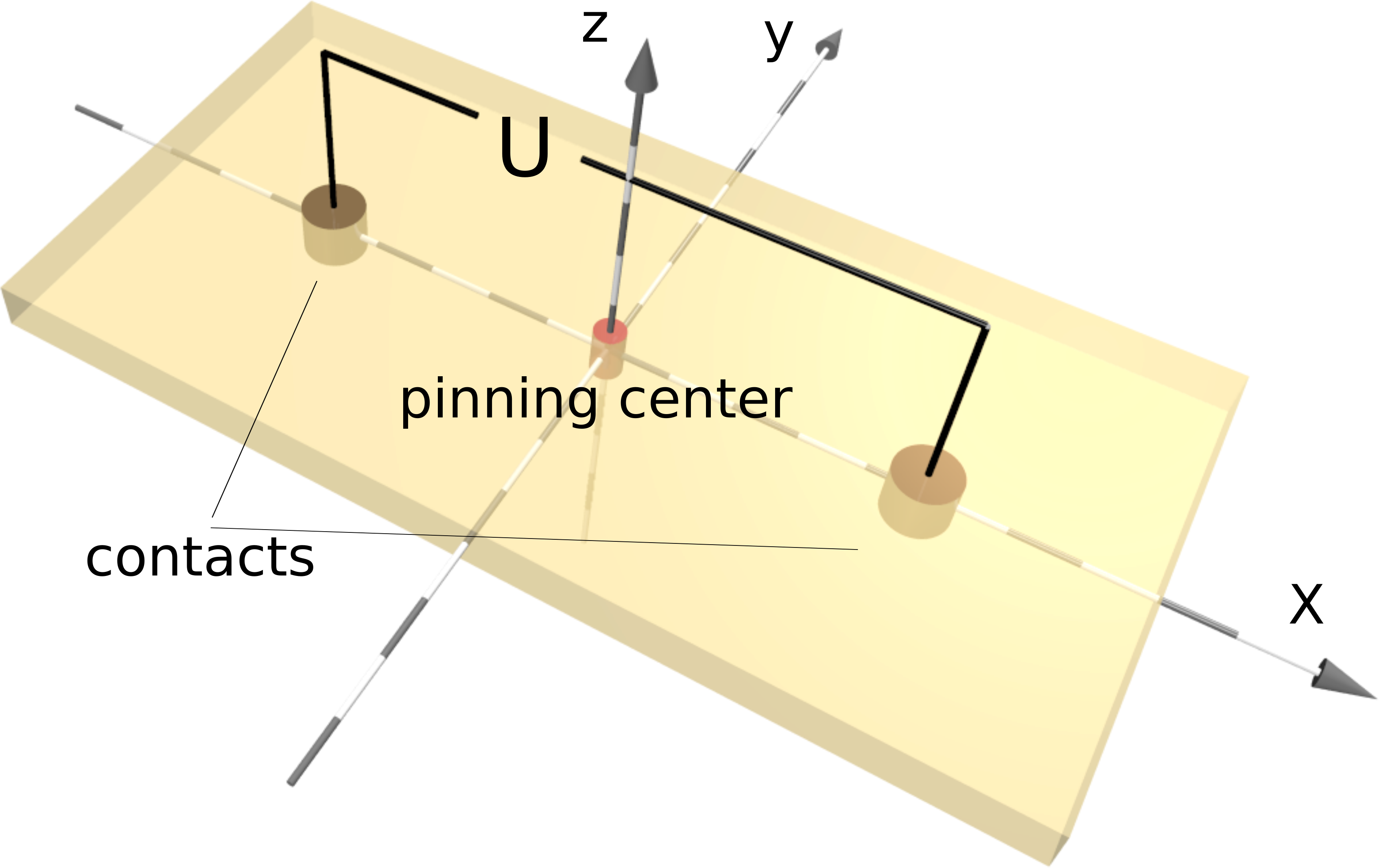}
   \caption{Schematic plot of the setup: A voltage $U$ is applied in between the two contacts of the ferromagnetic thin film. A potential pinning center is located in the center of the film.
}\label{setup}
  \end{figure}

We consider a metallic ferromagnetic thin film, see Fig.~\ref{setup},
with two metallic contacts embedded on opposite sides of the film through which we apply a voltage $U$. 
In the figure the coordinate system is chosen such that the magnetic film is placed in the x-y plane. The two contacts are located symmetrically around the center of the ribbon. 
We prepare an initial state in which a magnetic skyrmion is stabilized by the Dzyaloshinsii-Moriya interaction\cite{Dzyaloshinsky1958,Moriya1960b} (DMI) in the center.

\subsection{Physical picture}

The significant key ingredient of an RC system are the non-linear elements which function as network nodes. 
The non-linearity in a skyrmion network is obtained by exploiting spin-torque effects together with magneto-resistive effects. 
To be more concrete, we consider here a single skyrmion which is initialized at the center of the ferromagnetic ribbon and eventually pinned. Then we apply a voltage in between the two contacts which leads to a characteristic current flow that we compute based on the AMR effect as shown in Ref.~\onlinecite{Kruger2011}.
The resistance of the ``device'' is a function of the type, shape, relative size and position of the skyrmion in the ribbon. 

A pinned skyrmion is depinned from the impurity above a certain critical current density due to spin transfer torque effects.\cite{Schulz2012, Iwasaki2013, Rosch2013,Mueller2015,Sitte2016, Everschor-Sitte2016}  
Below the corresponding voltage the spin torques will cause the skyrmion to deform.
The deformation in the pinned regime in turn will change the resistance of the device due to the AMR effect resulting in its non-linear current-voltage characteristics.

\subsection{Numerical model for spin dynamics and current flow}
Our numerical simulations are based on the Landau-Lifshitz-Gilbert (LLG) equation for the unit vector field $\vect{m}= \vect M/M_s$ with the magnetization $\vect M$ and the saturation magnetization $M_s$. The equation is generalized to include spin-transfer-torque effects induced by the electric current 
which is calculated self-consistently based on an AMR module\cite{Kruger2011} due to an applied voltage $U$ between the two contacts. The corresponding LLG equation takes the following form
\begin{equation}
(\partial_{t} + \xi\, \vect j[U,\vect m] \cdot \vect{\nabla} ) \vect{m} = -\gamma \vect{m} \times \vect{B}_{\mathrm{eff}} +\alpha \vect{m} \times (\partial_{t} + \frac{\beta}{\alpha} \xi\, \vect j[U,\vect m] \cdot \vect{\nabla}) \vect{m},
\end{equation}
where $\gamma$ is the gyromagnetic ratio, and $\alpha$ and $\beta$ are {the} dimensionless Gilbert {damping} and non-adiabatic {spin-transfer-torque} parameters, respectively.
The prefactor $\xi =P \mu_{B}/(e M_{s})$ is proportional to the current polarization $P$ with $\mu_{B}$ denoting the Bohr magneton and $e$ the electron charge. 
The effective magnetic field is given by $\vect{B}_{\mathrm{eff} }= -M_{s}^{-1} (\delta F[\vect{m}]/\delta \vect{m})$, where $F$ describes the free energy of the system 
\begin{align*}
F&= \int \left( A_{\mathrm{ex}} (\nabla \vect{m})^{2} + K_{u} (1-m_z^2)
- \frac{\mu_0}{2} M_S \vect m \cdot \vect H_d(\vect m) \right) dV 
+F^{\mathrm{B}/\mathrm{N}}_{\mathrm{DMI}}[\vect m], \quad \text{with}\\
F^\mathrm{B}_{\mathrm{DMI}}&= \int  D_B \vect{m}\cdot (\nabla \times \vect{m})\, dV,\\
F^\mathrm{N}_{\mathrm{DMI}}&= \int D_N \vect{m} \cdot \lbrack(\mathbf{z}\times \nabla)\times \vect{m}\rbrack dV,
\end{align*}
where $A_{\mathrm{ex}}$ is the exchange constant, $K_{u}$ is the easy-axis anisotropy strength, and
the third term describes the dipolar interactions.
$F^{\mathrm{B}/\mathrm{N}}_{\mathrm{DMI}}$ denotes the Bloch / N\'eel DMI.\cite{Dzyaloshinskii1957, Moriya1960b, Thiaville2012}
The relaxation of the current happens on a short time scale compared to the dynamics of the magnetization. Therefore, the current $j[U,\vect m]$ can be calculated self-consistently based on the AMR effect\cite{Kruger2011} 
where the current density depends on the magnetization through the conductivity tensor $\vect \sigma[\vect m]$ with \begin{equation}
\label{j}
\vect j[U, \vect m] = -\vect \sigma[\vect m] \cdot \vect E[U].
\end{equation}
The electric field $\vect E$ is generated by the applied voltage $U$ and the corresponding potential is given by $\vect E=-\nabla \Phi$, 
where $\Phi$ satisfies the Poisson equation with the boundary conditions \mbox{$\Phi|_{c1}=-\Phi|_{c2}=U$} at the two contacts. 
The conductivity tensor is deduced from the inverse of the resistivity tensor $\vect \rho[\vect m]$.
Assuming that the resistivity consists of a material-dependent isotropic part and an anisotropic part that is given by the AMR, the resistivity is a function of the angle $\theta$ between current and magnetization direction. 
The local total resistivity is then given by 
\begin{align}\label{amr}
\rho_{tot}=\rho_{\perp} + (\rho_{\parallel}-\rho_{\perp})\cos^2 \theta, 
\end{align}
where $\rho_{\parallel}$ and $\rho_{\perp}$ denote the resistivities for currents flowing parallel and perpendicular to the magnetization direction. 
Since the scalar projection of the magnetization in the direction of the current is given by $\cos{\theta}$, the corresponding tensor takes the form 
$\vect \rho[{\vect m}]=\rho_{\perp} \mathbb{1}+ (\rho_{\parallel}-\rho_{\perp}) \hat{P}[\vect m]$, where the projection operator is given by $\hat{P}[\vect m]=\vect m \otimes \vect m$.
Thus, for the conductivity tensor it follows,
\begin{align} \label{sigma}
\begin{split}
\vect \sigma[\vect m] &= \frac{1}{\rho_{\perp}} \mathbb{1} + \left( \frac{1}{\rho_{\parallel}} - \frac{1}{\rho_{\perp}} \right) \hat{P}\vect[\vect m] \\
&= \sigma_0 \left( \mathbb{1} - \frac{6a}{6+a}
 \begin{pmatrix}
 m_x^2 - \frac{1}{3} & m_xm_y & m_x m_z \\
 m_ym_x & m_y^2 -\frac{1}{3} & m_ym_z \\
 m_zm_x & m_zm_y & m_z^2-\frac{1}{3}
 \end{pmatrix}\right),
 \end{split}
\end{align}
with $\sigma_0= (1/\rho_{\parallel} + 2/\rho_{\perp})/3$ and the AMR ratio $a=\frac{2(\rho_{\parallel}-\rho_{\perp})}{\rho_{\parallel} + \rho_{\perp}}$ as shown in Ref.~\onlinecite{Kruger2011}.

\section{Current distribution for a single skyrmion based on the AMR effect}
\label{results1}

In this section we study the current distribution based on the AMR effect for Bloch and N\'eel skyrmions due to the applied voltage.

\subsection{Current profile}

\begin{figure}
\centering
\includegraphics[width=1\textwidth]{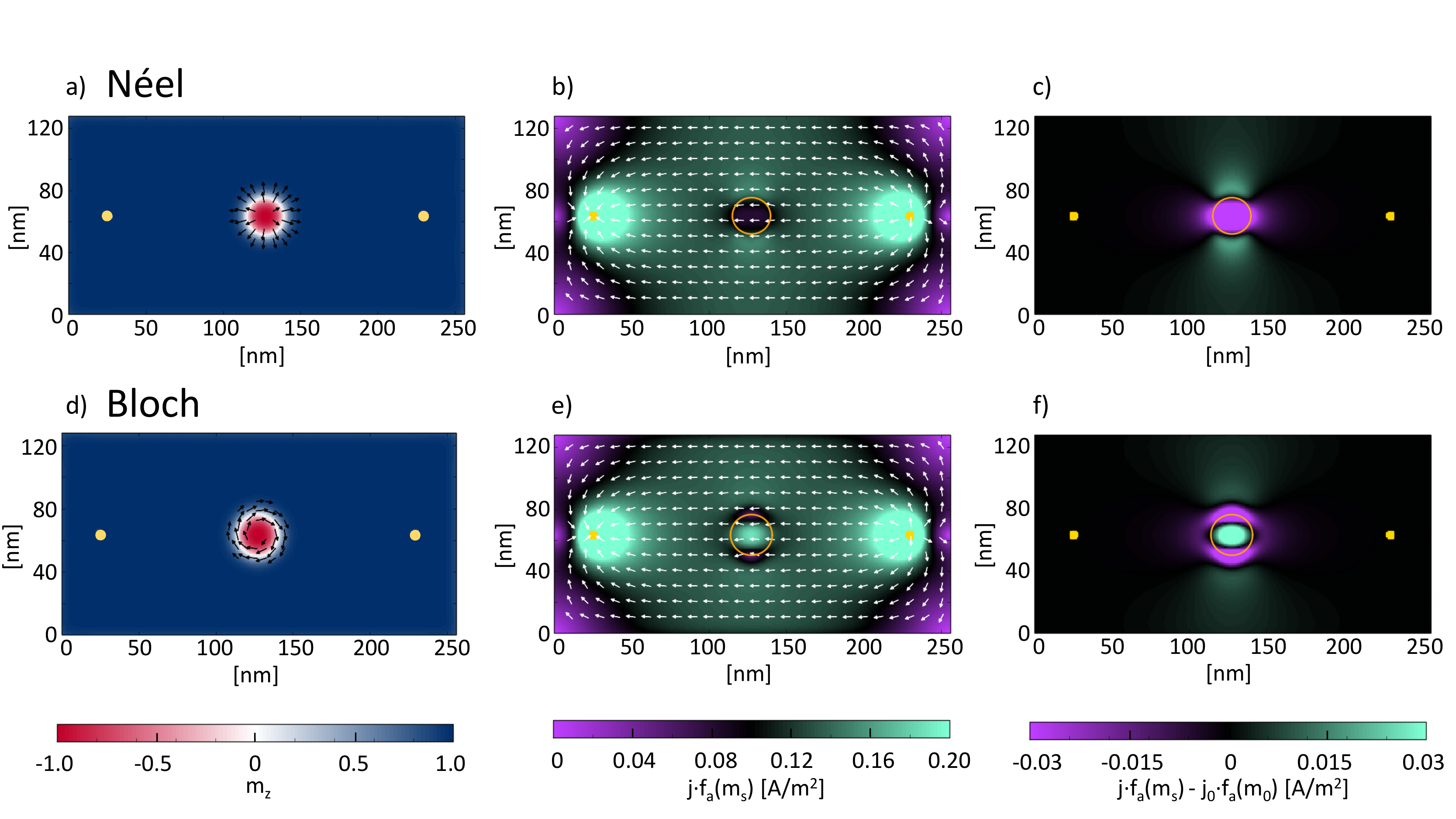}
\caption{
Magnetization and current profiles for N\'eel and Bloch skyrmion shortly after applying a voltage in-between the contacts (yellow dots).
In a) and d) we show the magnetization configuration of a  N\'eel and a Bloch skyrmion, respectively. The color code denotes the out-of-plane component and the black arrows indicate the in-plane components. 
b) and e) show the corresponding current flow due to an applied voltage. Here the color code indicates the current strength and the white arrows refer to the direction of the current. 
c) and f) reveal the effect of the skyrmions after subtracting the current density of the trivial out-of-plane ferromagnetic state. 
The circle in the current plots indicates the position of the skyrmion defined by the region where the out-of-plane magnetization vanishes.
As can be seen from the plots, the current profiles differ for both skyrmion types. 
For the N\'eel skyrmion the current is expelled from the center and prefers to flow in $x$ direction tangential to the circle
whereas for the Bloch skyrmion the current predominantly flows through the center of the skyrmion.
The parameters used for these simulations are: $\alpha = 0.25$, 
$M_s=\SI{4.9e5}{\ampere\per\metre}$, 
$\sigma_0=\SI{5e6}{\siemens\per\metre}$, 
$U = \SI{3.5e-2}{\volt}$, 
$A_{ex}=\SI{6e-12}{\joule\per\metre}$, 
$K_u=\SI{3e5}{\joule\per\cubic\metre}$, 
and $D_{B/N}=\SI{1e-3}{\joule\per\square\metre}$.
These parameters are similar to the ones used in simulations modeling real materials exhibiting skyrmion physics.\cite{Moutafis2015,Woo2016}
 The value for the bare resistance for this ribbon is given by $R_0 =\SI{348.4}{\ohm}$ for the N\'eel-type and $R_0 = \SI{350.6}{\ohm}$ for the Bloch-type skyrmion. The current densities are scaled with respect to the AMR ratio, $a$. The scaling function, $f_a$, increases with $a$ and changes with the local magnetization components. It is constraint by $f_0=0$ and $f_1=1$. Here, $m_s$ and $m_0$ refers to the magnetization profiles with and without a skyrmion, respectively. For more details see Sec.~\ref{sec:micromagsim}.}
  \label{fig:path}
 \end{figure}

The current profiles for Bloch and N\'eel skyrmions based on the AMR effect are different.
Representative results for the magnetic textures and the corresponding current paths are shown in Fig.~\ref{fig:path}a,b,d,e. 
To illustrate the bare effect of the skyrmions, we also substract the current path obtained for the trivial out-of-plane ferromagnetic state, see Fig.~\ref{fig:path}c and Fig.~\ref{fig:path}f.

The main result is that for the N\'eel skyrmion (upper row) the current is expelled from the center of the skyrmion.
In contrast, the current density is enhanced in the center for the Bloch skyrmion. 
For details of the numerics see Sec.~\ref{sec:micromagsim}.

The resistance of the device not only depends on the type, shape and relative size of the skyrmion with respect to the device, but also on its position due to a non-uniform current density. 
Therefore a motion of the skyrmion will change the resistance of the device. We first focus on the change of the resistance when the skyrmion is moving. For this purpose, we study the trajectory of the unpinned skyrmions due to an applied voltage and its resulting current flow.

\subsection{Trajectories of skyrmions}
\begin{figure}
 \includegraphics[width=1\textwidth]{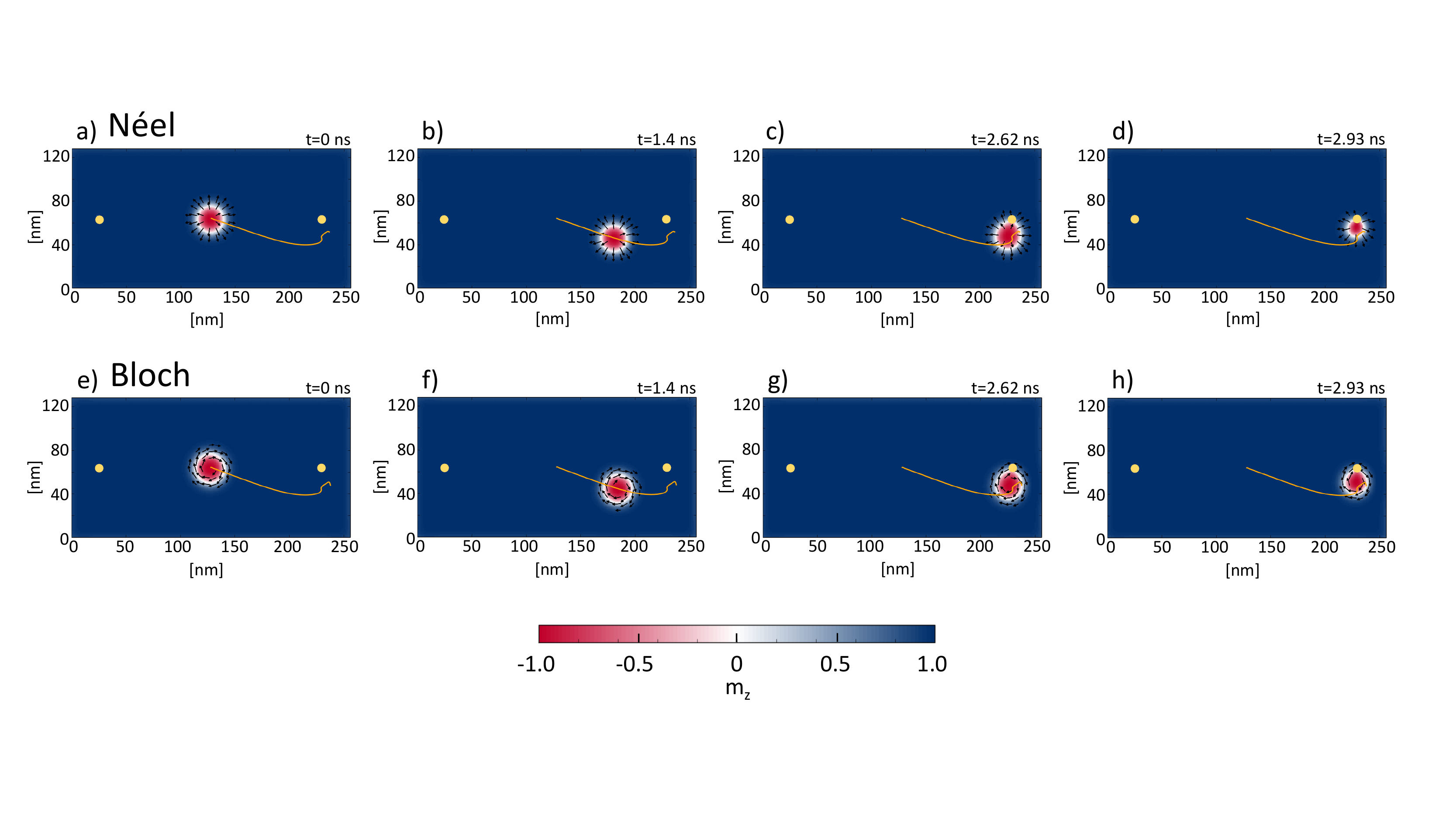}
 \caption{Evolution in time of N\'eel (top) and Bloch-type (bottom) skyrmions for a fixed voltage. The magnetization profile is encoded in analogy to Fig.~\ref{fig:path} a) and d). 
The parameters used for the simulations are the same as in Fig.~\ref{fig:path}, so that a) and e) of this Figure correspond to Fig.~\ref{fig:path}c and Fig.~\ref{fig:path}f, respectively. During this motion a significant deformation of the skyrmion is only observed in the vicinity of the contact.}
  \label{fig:motionpath}
 \end{figure}
\begin{figure}
 \includegraphics[width=1.\textwidth]{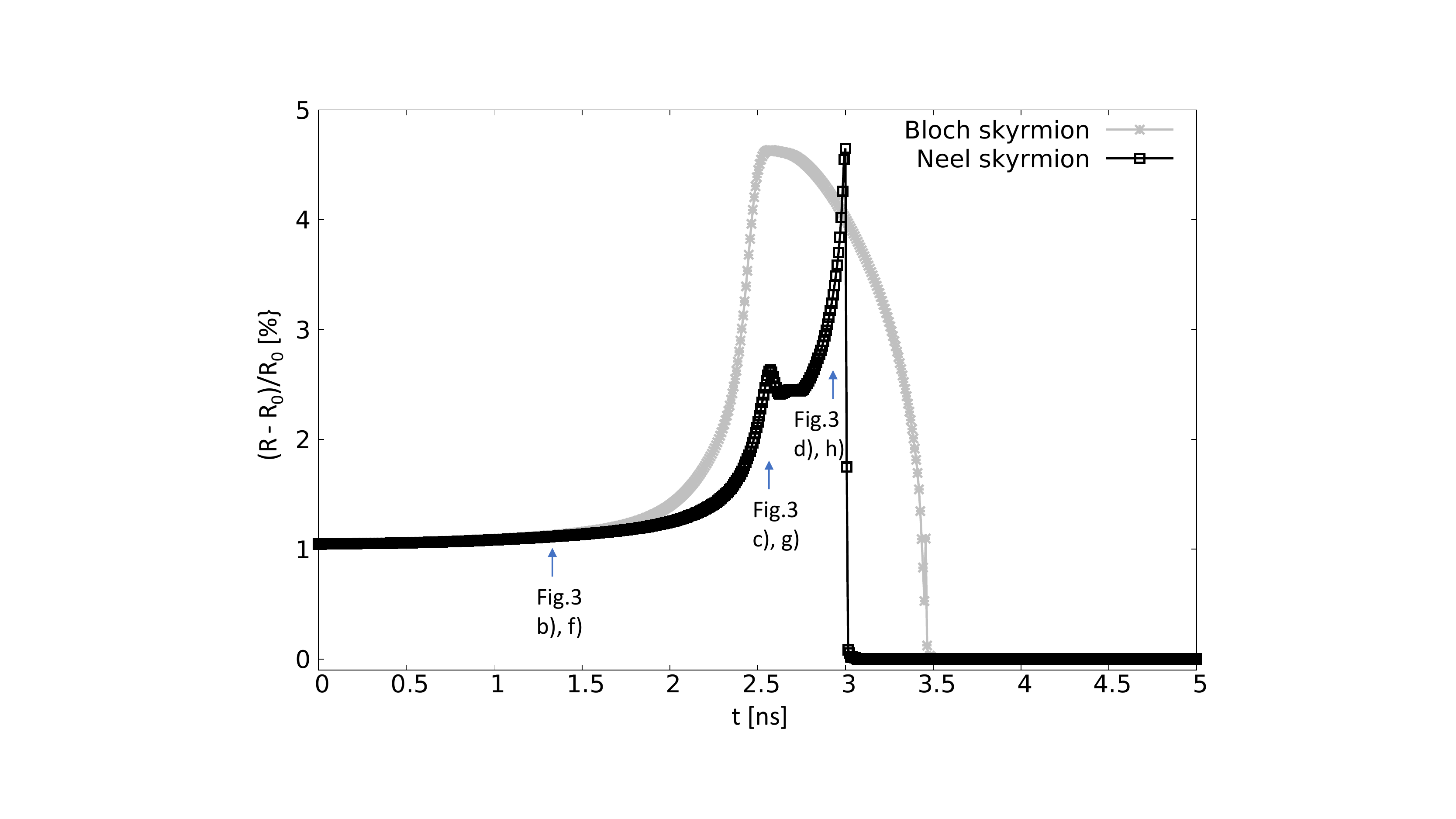}
 \caption{Relative resistance dependence $(R-R_0)/R_0$ on time for a N\'eel-type and Bloch-type skyrmion at a fixed voltage. Here, $R_0$ denotes the resistance of the out-of-plane ferromagnetic configuration in the absence of a skyrmion.
The resistance mainly increases when the skyrmion approaches the contact until it is absorbed. Details of the curves depend on the magnetization texture, see discussion in the main text. 
The annotations refer to the magnetization profiles of the skyrmions at the corresponding times.
The parameters used for the simulations are the same as in Figs.~\ref{fig:path}  and \ref{fig:motionpath}.
The value for the bare resistance in this geometry is given by $R_0 =\SI{348.4}{\ohm}$ for the N\'eel-type and $R_0 = \SI{350.6}{\ohm}$ for the Bloch-type skyrmion.}
 \label{fig:Rt}
 \end{figure}

After applying a voltage, the skyrmions move due to spin-transfer-torque effects. Representative skyrmion trajectories are shown in Fig.~\ref{fig:motionpath}. 
Here, the applied voltage is small enough to not noticeably modify the shape of the skyrmion. 
Since the total resistance of the device does depend on the position of the skyrmion, this setup allows for a time-dependent resistance variation at a fixed applied voltage. 
The relative resistance as a function of time, see Fig.~\ref{fig:Rt}, 
shows how the resistance changes with time when the skyrmion is approaching one of the contacts, until it eventually disappears. 

The details of the time evolution of the magnetic texture and the corresponding resistances depend on the microscopic parameters of the device. 
In general, the resistance increases when the skyrmion approaches the contact since here the current density is largest. The contribution to the AMR becomes larger because the resistance is determined by the interplay between the local current density and the magnetization profile, see Eqs.~\eqref{j} - \eqref{sigma}. 
Depending on the applied voltage strength a skyrmion merges with the contact or passes it. 
For the parameter set used in Figs.~\ref{fig:path}, \ref{fig:motionpath} and \ref{fig:Rt}, the skyrmion disappears at the contact. 
Here, in the vicinity of the contact the two skyrmion types show a different behavior. 
The resistance of the N\'eel-type skyrmion shows two peaks which can be most intuitively understood when looking at the cross section of the skyrmion. 
Its cycloidal domain wall profile contains two points where the out-of-plane magnetization component is zero, denoted as $r_1$ and $r_2$ in the following.
The first local maximum in the resistance appears when the first point, $r_1$, with a large parallel contribution to AMR touches and eventually merges with the contact.
Subsequently, the second point $r_2$ is moving towards the contact. As it also causes a larger resistance, this explains the second maximum.
The Bloch-type skyrmion only shows one peak, namely when the skyrmion touches the contact. Since the main contribution to AMR does not come from the connecting line of the contact and the center of the skyrmion, we do not observe an increase in the resistance when the skyrmion starts to shrink.

\section{Non-linear I-V characteristics from local anisotropy}
\label{results2}

\begin{figure}
  \includegraphics[width=1\textwidth]{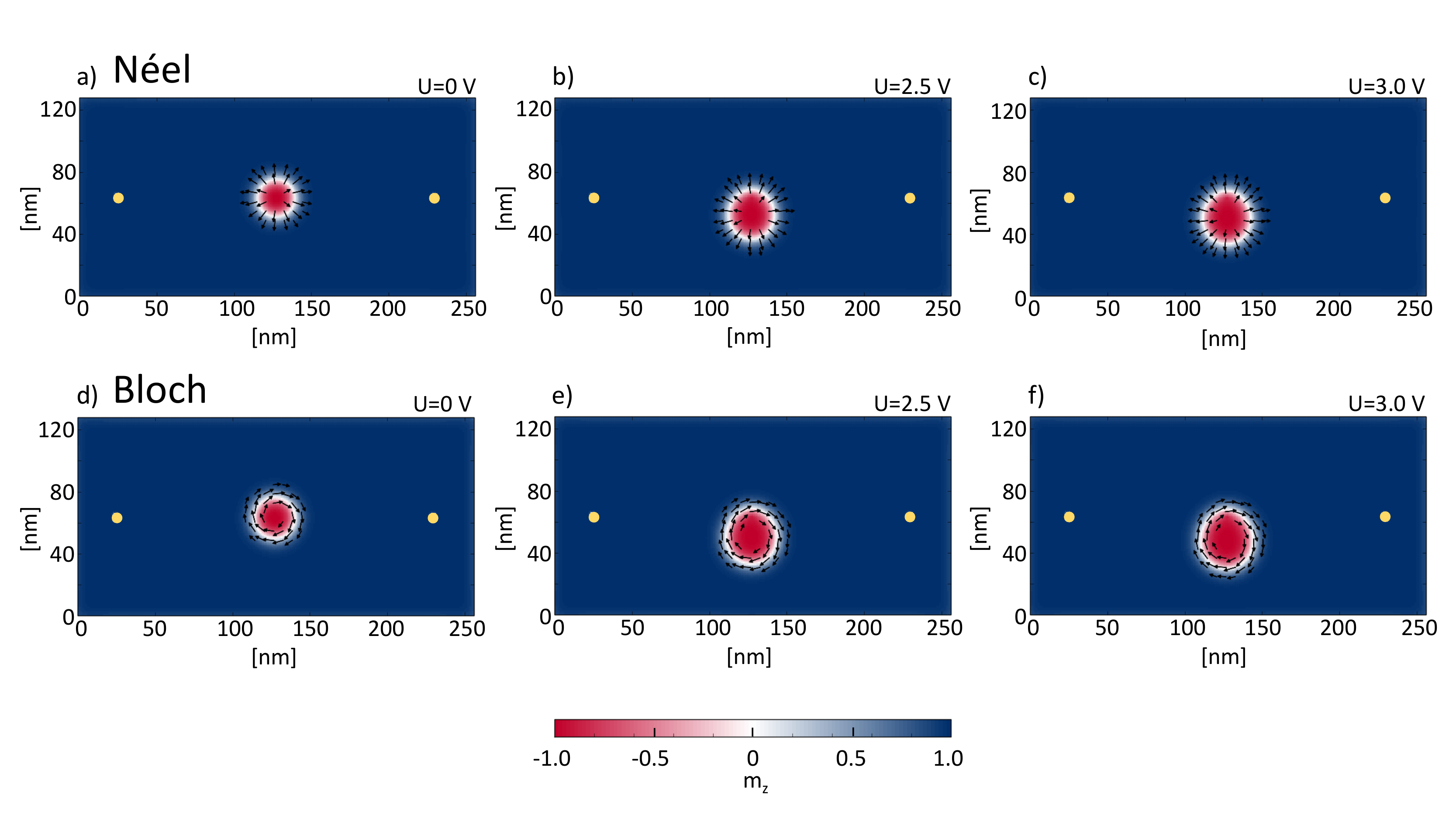}
  \caption{%
  Deformation of a pinned N\'eel (top) and Bloch type (bottom) skyrmion due to an applied voltage 
  ($U_0=\SI{0}{\volt}$, $U_1=\SI{2.5e-2}{\volt}$ and $U_2=\SI{3.0e-2}{\volt}$) which gives rise to a non-uniform current density. With increasing voltage the deformation increases.
  The radius $r_p$ and the  
  local anisotropy strength $K_p$ of the pinning center are $r_p=\SI{1.5}{\nano\metre}$ and
  $K_p= \SI{3e6}{\joule\per\cubic\metre}$, respectively. 
  The remaining parameters have been chosen as follows: $A_{ex}=\SI{6e-12}{\joule\per\metre}$,
  $K_u=\SI{3e5}{\joule\per\cubic\metre}$, and $D_{B/N}=\SI{1e-3}{\joule\per\square\metre}$.}
\label{fig:deforming}
\end{figure}

In this section we consider the situation where the skyrmion is pinned at the center. 
We model the pinning center by a small cylindrically shaped area in the center of the ribbon with an enhanced out-of-plane anisotropy strength. 

\begin{figure} 
  \includegraphics[width=1.0\textwidth]{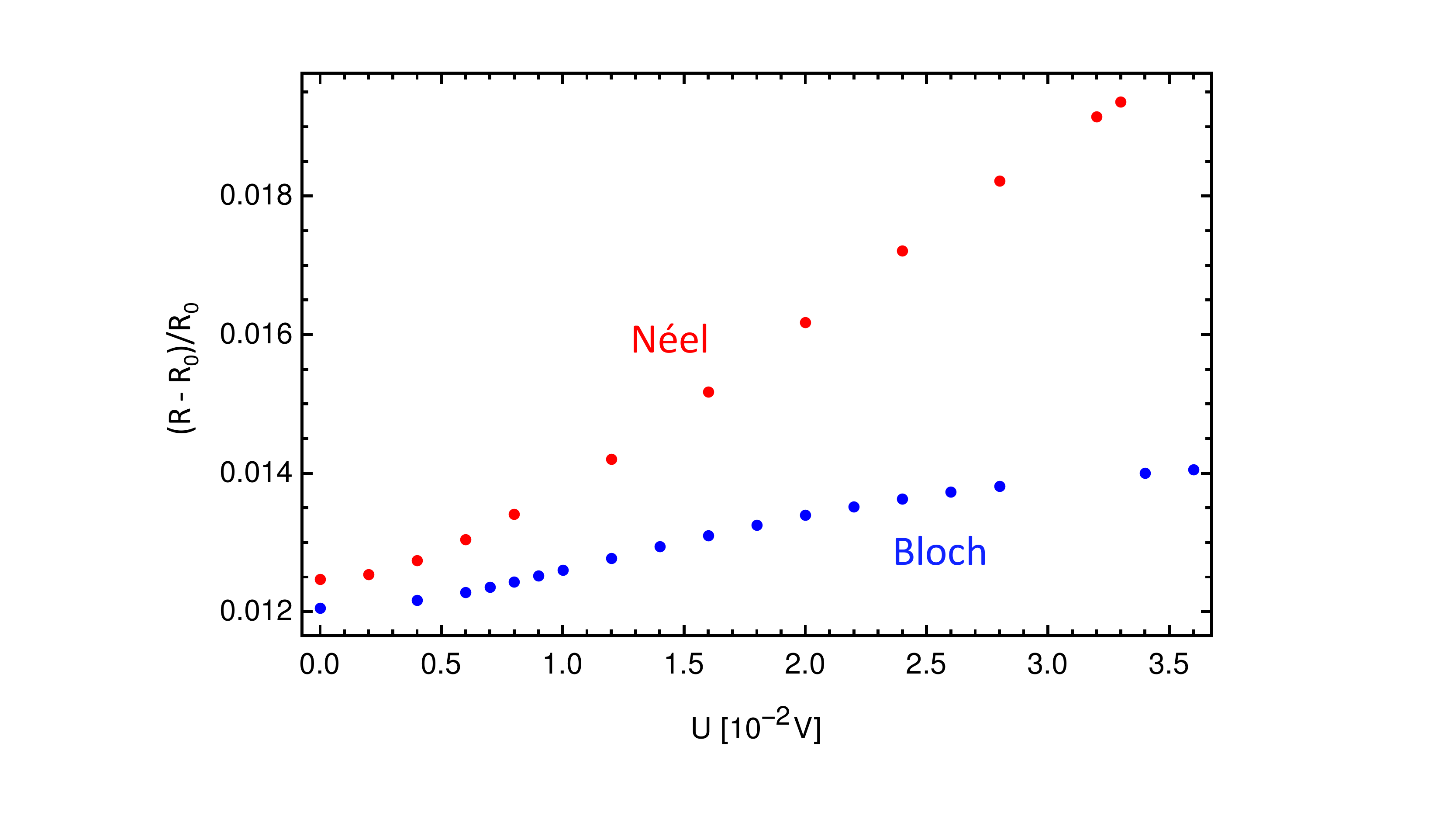}
  \caption{Relative resistance contribution of a pinned skyrmion as function of an applied voltage
  visualizing the non-linear behavior of the device.
  The parameters used are given in Fig.~\ref{fig:deforming}. }
   \label{fig:UR}
 \end{figure}

The interplay of pinning and spin-transfer-torques causes the skyrmion to deform/stretch as long as the applied voltage does not exceed a critical value. The deformation is stronger, the larger the applied voltage is, as shown in Fig.~\ref{fig:deforming}.\cite{Sitte2016, Everschor-Sitte2016}
Since the resistance is a function of the local angle between the current and the local magnetization, it varies with the texture of the skyrmion and therefore becomes a function of the applied voltage $U$. 
Thus, the device has a non-linear current-voltage characteristics with the current-voltage relation 
\begin{align}
 \frac{dI}{dU}=G(U,\vect m) = R^{-1} (U, \vect m).
\end{align}
Here $G$ ($R$) denotes the differential conductivity (resistance) function depending on the applied voltage $U$ and the magnetic texture $\vect m$.
The corresponding behavior of the relative resistance of the device as a function of the applied voltage is plotted in Fig.~\ref{fig:UR} demonstrating its non-linear current-voltage characteristics.
 
\begin{figure}
 \includegraphics[width=1\textwidth]{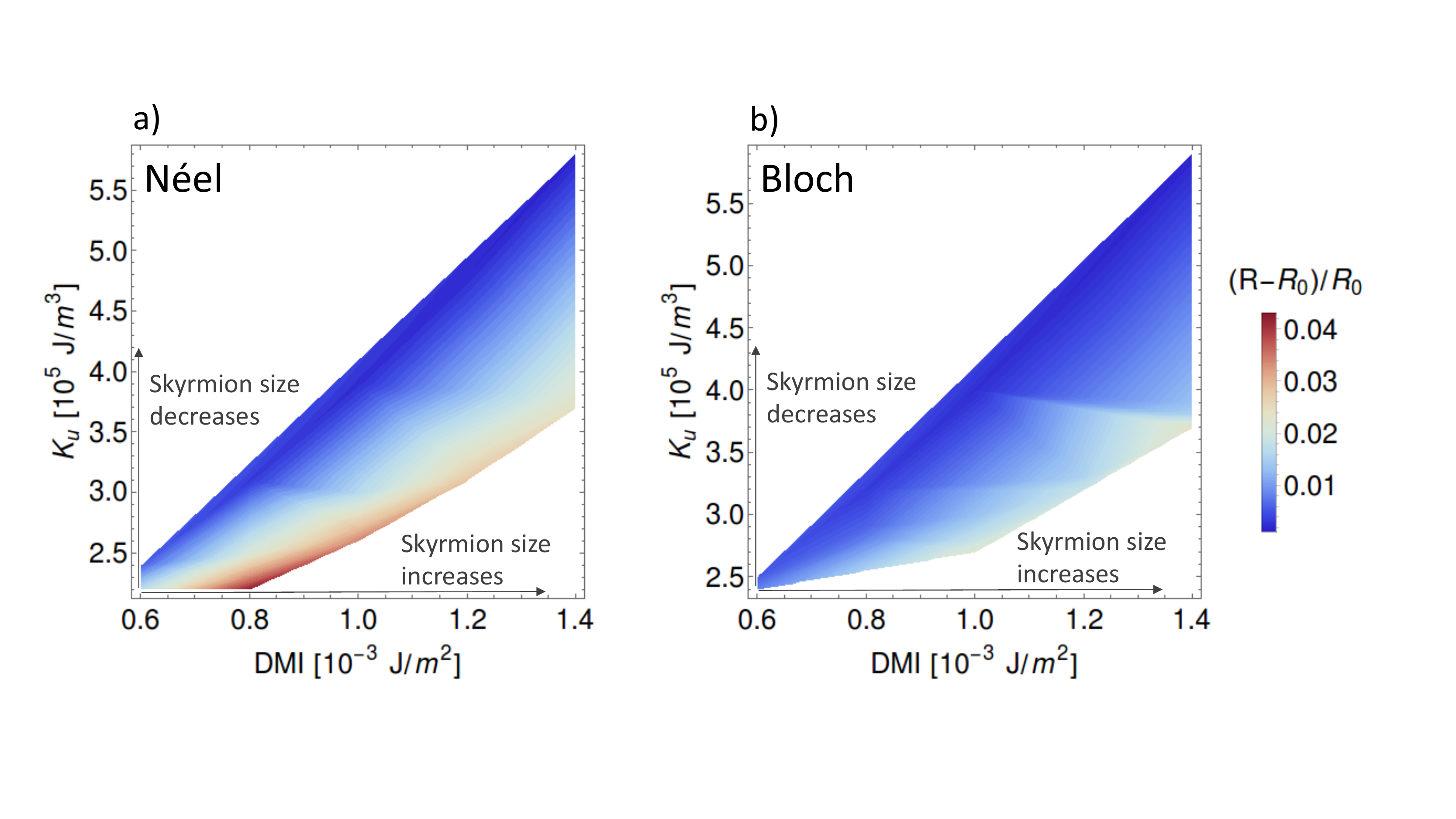}
 \caption{Relative resistance as a function of DMI and anisotropy for a N\'eel (left) and Bloch-type (right) skyrmion at a fixed voltage, $U=\SI{3.0e-2}{\volt}$. The magnitude of the resistance rises with increasing skyrmion size being determined by the interplay between DMI and anisotropy strength.   
For the value of exchange coupling and the pinning center we have used the same parameters as in Fig.~\ref{fig:deforming} and Fig.~\ref{fig:UR}.
 }
  \label{fig:phasediag}
 \end{figure}

In Fig.~\ref{fig:phasediag} we varying the DMI and the anisotropy strength and analyze a range in the parameter set for which we obtain skyrmion configurations that remain pinned when a voltage of $U= \SI{3.0e-2}{\volt}$ is applied. The color in this plot encodes the magnitude of the change in resistance.
In general we conclude that a larger skyrmion leads to a larger effect.
For our finite-size device, the skyrmion size shrinks when the anisotropy strength is increased 
whereas it increases with larger DMI strength. The latter causes the canting of the local spins to be extended more into the ferromagnetic strip whereas the domain wall itself is kept narrow. 

In the phase diagram, the AMR effect has been studied as a function of the anisotropy and the DMI strength which both affect the size of the skyrmion. However, the resistance also changes with the system width which emphasizes that the relative size of the skyrmion with respect to the ribbon plays a crucial role for AMR. 
The change of the system width is associated with the interplay of two effects which first leads to an increase and later to a reduction of the resistance with decreasing ribbon width. 
The resistance increases with a smaller system size since the effect of anisotropy and exchange interaction relative to the DMI is reduced. This results in a larger AMR effect.  
However, below a critical width the resistance for a Bloch-type skyrmion starts to decrease again. With decreasing system width more current gets pushed from the boundaries towards the center where for a Bloch-type skyrmion the resistance is lowest. 

Within this section we have analyzed the AMR based behavior of a single skyrmion confined to a two dimensional ferromagnetic device. 
We have demonstrated its non-linear current-voltage characteristics making it a promising candidate as a basic element for a reservoir computing network.

\section{The path to reservoir computing with skyrmion fabrics}
\label{sec:pathtoRCwithskrymions}

\begin{figure}
         \includegraphics[width=1\textwidth]{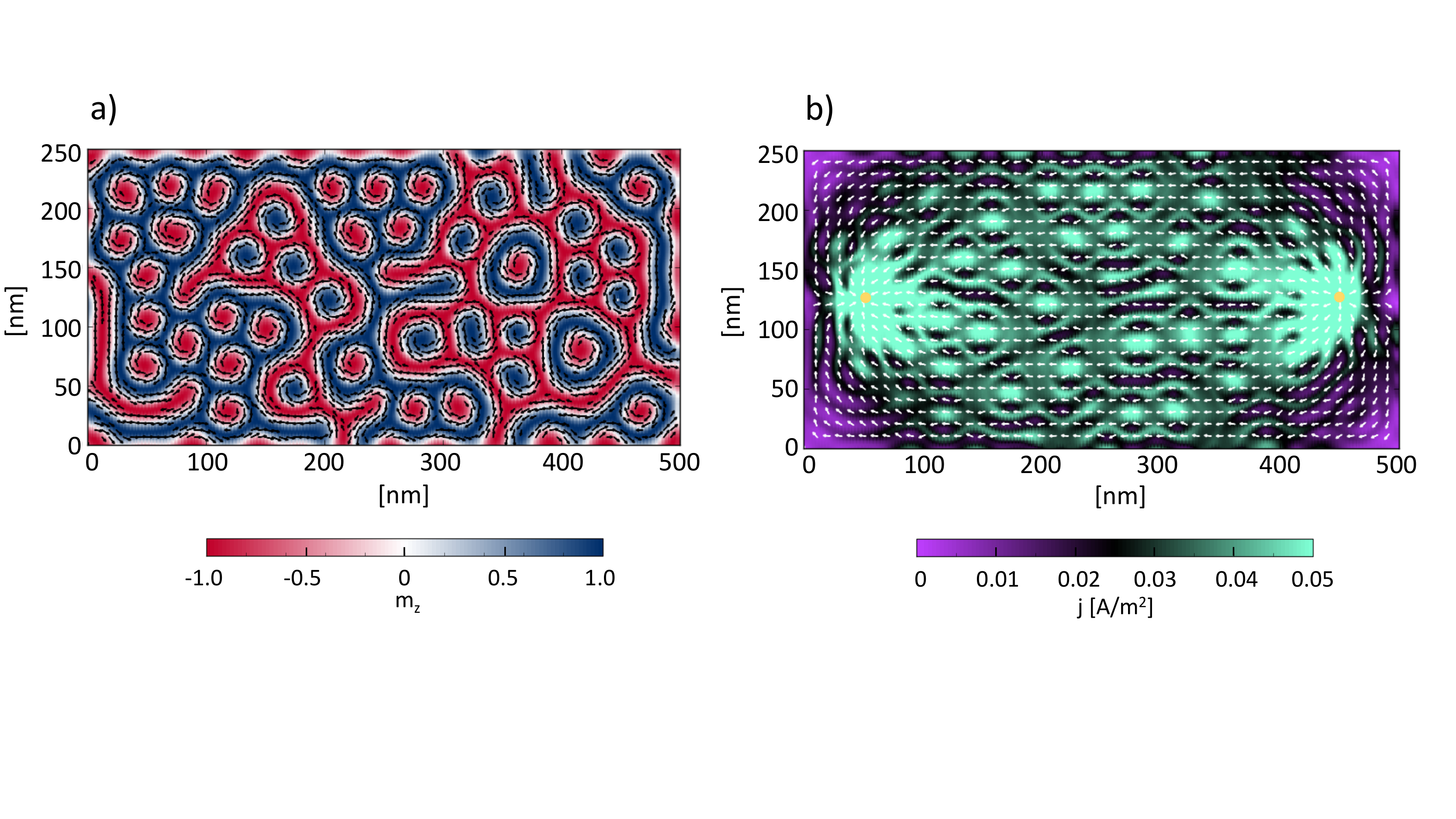}
  \caption{Magnetization and current profile for frustrated magnetic texture including Bloch skyrmions after applying a voltage in-between the contacts. a) shows the contour plot of the out-of-plane magnetic configuration. The blue and red droplets correspond to skyrmions.  The white lines are domain walls separating the skyrmions. The arrows indicate the in-plane magnetization. In b), the current path is shown. The parameters used for these simulations are: $\alpha = 0.5$, 
$M_s=\SI{4.9e5}{\ampere\per\metre}$, 
$\sigma_0=\SI{5e7}{\siemens\per\metre}$, 
$U = \SI{1.0e-3}{\volt}$, 
$A_{ex}=\SI{6e-12}{\joule\per\metre}$, 
and $D_{B}=\SI{2e-3}{\joule\per\square\metre}$.
For more details see Sec.~\ref{sec:micromagsim}.}
 \label{fig:skyrmionRC}
\end{figure}

So far we have discussed the device characteristics of a single skyrmion in a conducting ferro-magnetic wire. 
To be useful for information processing applications, these single skyrmions must be connected and arranged into a random, self-assembled, recursively connected configuration. 
Self-organization can be achieved in a ferromagnetic substrate where, for example, DMI competes against the exchange interaction leading to a frustrated order configuration as shown in Fig.~\ref{fig:skyrmionRC}a, exemplified for Bloch DMI.
A current flowing through the magnetic substrate follows the path of least resistance. 
As imposed by the AMR effect, the current flows between the domain walls as shown in Fig.~\ref{fig:skyrmionRC}b.  

Thin films of self-organized skyrmions connected by current carrying domain walls bear at least a superficial resemblance to cortical networks of neurons, axons, synapses and dendrites.
In particular, the above illustration suggests an approach for implementation of RC systems with skyrmion-based magnetic substrates. 
There is a correspondence between the spin configuration of the magnetic layer shown in Fig.~\ref{fig:skyrmionRC}a and atomic switch networks studied in Refs.~\onlinecite{Hasegawa2010, TingChang2013, Sillin2013}. 
Here, individual skyrmions correspond to the silver sulfide (AgS) atomic switches and the low-resistance parts of the domain walls connecting the skyrmions correspond to silver (Ag) nanowires. 
In this analogy, the non-linear current-voltage relationship of (pinned) skyrmion fabrics would be used instead of the $\sinh$ non-linear current-voltage relation which characterizes AgS atomic switches. 
Based on the configurational resemblance between skyrmion networks and atomic switch networks and the similar non-linear current-voltage characteristics, we envisage that magnetic skyrmion substrates will be a suitable implementation for RC models.  Additional analysis will be needed to establish a firm link between this novel combination of a computational methodology to a physical implementation. 

Relative to atomic switch networks, skyrmion networks offer several potential advantages for RC implementation.
As stated above, skyrmions have been observed as small as \SI{1}{\nano\metre}, and \SI{10}{\nano\metre} to \SI{50}{\nano\metre} diameter is quite common, whereas the elements of atomic switch networks are a factor of 10 to 100 larger.
Due to the efficient coupling between conduction electrons and skyrmions even low current densities influence the magnetic texture.  
A rough estimate comparing the power consumption of atomic switch networks and skyrmion fabrics shows that skyrmion systems require about two orders of magnitude lower power.
Standard voltages used for atomic switch networks are of the order of \SI{1}{\volt} and resistances are about \SI{10}{\kilo\ohm}\cite{Avizienis2012, Stieg2012, Sillin2013}
 leading to a power of $P \approx \SI{100}{\micro\watt}$. Whereas for our skyrmion system we obtain powers of the order of $P \approx \SI{1}{\micro\watt}$.
In general, skyrmion systems interacting with currents and magnons have many internal degrees of freedom. 
This complexity results in reservoirs with high tunability and functionality much larger than for memristors and atomic switch networks. 
As shown in Fig.~\ref{fig:skyrmionRC}, a skyrmions tend to cluster into ensembles of various sizes where each cluster is surrounded by a domain wall. The details of the clustering depend on the competing micromagnetic interaction strengths, therefore it also should be very sensitive to small bias fields. 
However, this analysis goes beyond the scope of this work. 
The present approach is similar to the resistive switch network but promises to
have advantages regarding size, energy efficiency, complexity and adaptivity as summarized in the following:
  \begin{enumerate}
  \item Energy efficiency:  The power requirements are potentially low due to an efficient coupling between skyrmions and currents.
  \item Size:  The size of reservoir elements is relatively small. Assuming an average skyrmion diameter of about \SI{10}{\nano\metre}, a million-element reservoir measures about \SI[product-units = repeat]{10 x 10}{\micro\metre}.
  \item Complexity: Skyrmion interactions with currents and magnons yield more internal degrees of freedom than a simple scalar resistivity associated with memristors and atomic switch networks.
  \item Adaptivity: High tuneability and adjustability of the network topology based on competing micromagnetic forces. 
  \item Homeostatic operation:  The system dynamics will be sensitive to  thermal effects or bias fields applied uniformly to the reservoir.  These can, in principle, be used to maintain homeostatic operating points.  
\end{enumerate}

\section{Summary and conclusions}

In this work we have proposed skyrmion systems for reservoir computing. 
To start with, we have shown a non-linear 
current-voltage characteristics for an isolated skyrmion pinned in a ferromagnetic strip based on the AMR effect. However, the particular type of magnetoresitive effect is not crucial for the consequences regarding the application of skyrmion networks for RC as other magnetoresistive effect also give rise to non-linear properties of such a device. The essential feature is the dependency of the current density on the skyrmion spin structure. 
Our results imply that skyrmion networks indeed might  provide a suitable physical implementation for RC applications.

\section{Micromagnetic simulations}
\label{sec:micromagsim}
The method we use is based on micromagnetic simulations which were performed using MicroMagnum\cite{MicroMagnum} including additional self-written software extension. As previously mentioned we simulate a quasi-two-dimensional ribbon.  For the results shown in Fig.~\ref{fig:path} to Fig.~\ref{fig:phasediag} the ribbon considered in the simulations is composed of \num{256 x 128} cells in the 
$x-y$-plane and a thickness of one cell in $z$-direction. The cells correspond to unit cubes of \SI{1}{\nano\metre} edge-length. In the center of the ribbon a cylinder along the $z$-direction is placed with an in-plane radius of \SI{15}{\nano\metre}. 
Two contacts are modeled as $Au$-cylinders close to the edges of the strip with a radius of \SI{3}{\nano\metre}, their centers are located at $10\%$ and $90\%$ of the ribbon along the $x$-direction and centered in $y$-direction. 
To mimic a pinning center we position a cylinder of \SI{1.5}{\nano\metre} radius with an enlarged anisotropy strength in the middle of the strip. \\
The initial magnetic profile corresponds to a ferromagnetic alignment in $z$-direction with a saturation magnetization of $M_s= \SI{4.9e5}{\ampere\per\metre}$ but with opposite magnetization of the ribbon and the cylinder.  The conductivity of the material is $\sigma_0 = \SI{5e6}{\siemens\per\metre}$ and its current polarization is $P=0.56$. Before the DC current is applied the system needs to be relaxed to the ground state. The single skrymion simulations are performed 
at a Gilbert damping parameter of $\alpha=0.25$ and a non-adiabatic spin transfer torque parameter of $\beta=0$. 

In contrast to the simulations with a single skyrmion, the ribbon in the $x-y$-plane used for the data in Fig.~\ref{fig:skyrmionRC} is composed of \num{500 x 250} cells of \SI[product-units = repeat]{1 x 1}{\nano\metre} lateral size. 
Here, to obtain the relaxed magnetization configuration we start from a completely random state with a saturation magnetization of $M_s= \SI{4.9e5}{\ampere\per\metre}$. The frustrated state is a result of the interplay between the exchange interaction and the DMI. The relaxation process is performed at a large Gilbert damping, $\alpha = 0.5$. The modeling of the contacts is similar to the one used for a single skyrmion but with a smaller contact radius of \SI{1}{\nano\metre}.  The material has the same conductivity. The spin polarization is $P=0.5$ and the ratio of the non-adiabatic and adiabatic spin-transfer-torque is $\frac{\beta}{\alpha}=0.02$. 

The AMR rate which enters the equation for the conductivity tensor and thus affects the magnitude of current density and resistance is set to $a=1$. The scaling function, $f_a$, is a function of the magnetization and has been derived from Eq.~(\ref{sigma}). For in-plane magnetization in $x$- or $y$-direction for instance it is given by $f_a = -7\frac{6+9a}{(6+a)}$. Depending on the position the factor for a ribbon with a skyrmion may differ from the one without a skyrmion.

\section*{Acknowledgments}

G. B. appreciates support and useful discussions with Narayan Srinivasa of Intel.
We acknowledge the funding from the Alexander von Humboldt Foundation,  the Transregional Collaborative Research Center (SFB/TRR) 173 Spin+X, the ERC Synergy Grant SC2 (No. 610115), and the Grant Agency of the Czech Republic grant no. 14-37427G. K.~E.-S. acknowledges funding
from the German Research Foundation (DFG) under the Project No. EV 196/2-1. D.~P. and K.~L. are recipients of a DFG-fellowship through the Excellence Initiative by the Graduate School Materials Science in Mainz (GSC 266) and K.~L. thanks the Studienstiftung des Deutschen Volkes for support.

\end{document}